\documentclass[preprint,prd,aps,showpacs,showkeys,nofootinbib]{revtex4}
\usepackage{epsfig}
\usepackage{graphicx}
\usepackage{amsmath}
\usepackage{amssymb}
\usepackage{booktabs}
\usepackage{xcolor}
\usepackage{float}
\usepackage{hyperref}
\UseRawInputEncoding

\begin{document}

	
	\title{Electroweak baryogenesis and electron EDM in the TNMSSM}
	
	\author{Xiang Yang$^{1,2,3}$\footnote{yangxiang202406@163.com}, Ming-Hui Guo$^{1,2}$, Jin-Lei Yang$^{1,2,3}$\footnote{jlyang@hbu.edu.cn}, Qing-Hua Li$^{1,2,3}$\footnote{lqhlqh202408.163.com}, Tai-Fu Feng$^{1,2,3,4}$\footnote{fengtf@hbu.edu.cn}}
	
	\affiliation{$^1$ Department of Physics, Hebei University, Baoding 071002, China}
	\affiliation{$^2$ Hebei Key Laboratory of High-precision Computation and Application of Quantum Field Theory, Baoding, 071002, China}
	\affiliation{$^3$ Hebei Research Center of the Basic Discipline for Computational Physics, Baoding, 071002, China}
	\date{\today}
	\affiliation{$^4$ Department of Physics, Chongqing University, Chongqing 401331, China}

\begin{abstract}
We have studied the impact of CP-violating (CPV) effects on electroweak baryogenesis (EWBG) and the electric dipole moment (EDM) of electron ($d_e$), the electric dipole moment (EDM) of mercury neutron ($d_n$) and the electric dipole moment (EDM) of ($d_{Hg}$) in an extension of the Minimal Supersymmetric Standard Model (MSSM). The model incorporating a $SU(2)$ triplet with hypercharges of $\pm1$ and a gauge singlet from the standard model that is mutually coupled, is collectively referred to as the next-to-minimal supersymmetric standard model with triplets (TNMSSM). Furthermore, we discuss the strong first-order phase transition (PT) achieved by this model via tree-level effects. The numerical results indicate that the TNMSSM can account for the observed baryon asymmetry. Additionally, the regions favored by EWBG can be compatible with the corresponding electron EDM bounds when different contributions to the $d_e$ cancel each other out.
\end{abstract}


\maketitle

\section{Introduction\label{sec1}}

Although the standard model (SM) has been experimentally confirmed in many theoretical predictions, there still exist phenomena that cannot be explained by the SM. One of the most intriguing issues among them is the baryon asymmetry of the universe (BAU)~\cite{Cooke:2013cba,Ade:2015xua}:
\begin{eqnarray}
	&&Y_B\equiv\frac{\rho_B}{s}=\left\{\begin{array}{l}
		(8.2-9.4)\times10^{-11}\;\;\;(95\%{\rm CL}),\;\;{\rm Big\;Bang\;Nucleosynthesis}\\
		8.65\pm0.09\times10^{-11},\qquad\quad\;\;\;\;\;\;\;\;{\rm PLANCK}\\
	\end{array}\right.\;
\end{eqnarray}
	where $\rho_B$ is the baryon number density, and $s$ is the entropy density of the universe. The matter-antimatter asymmetry provided by the SM's CP-violating (CPV) is not of the required magnitude, this indicates the presence of new physics (NP) beyond the SM. Explaining baryon asymmetry by electroweak baryogenesis (EWBG) necessitates new CPV effects to theoretically amplify the EWBG responsible for generating the matter-antimatter asymmetry~\cite{Kuzmin:1985mm}, which is also one of the most intriguing explanations for the origin of the asymmetry between matter and antimatter in cosmology.

The new CPV phases can not only provide many orders of magnitude higher matter-antimatter asymmetry than the SM predicts, but also yield larger values of the EDM than those predicted by the SM. In the SM, the predicted value of the electron EDM is approximately $10^{-38}{\rm e\cdot cm}$~\cite{Gavela:1981sk,Bernreuther:1990jx,Pospelov:2013sca}, yet it has not been detected by current experimental methods.
However, when new CPV phases are introduced, the electron EDM can be enhanced by several orders of magnitude, and it is quite possible to be dected in the near future. Such observations would provide compelling evidence for the existence of NP beyond the SM.
The upper bounds on the electron EDM $d_e$ have been presented in Refs.~\cite{ParticleDataGroup2024,Roussy2023}
\begin{eqnarray}
	&&|d_e|<4.1\times10^{-30}{\rm e\cdot cm}.
\end{eqnarray}
In addition, the neutron EDM $d_n$ and the mercury EDM $d_{Hg}$ will also impose strict constraints on the parameter space of the new model, with the relevant experimental data provided by Ref.~\cite{ParticleDataGroup2024}
\begin{eqnarray}
	&&|d_{Hg}|<7.4\times10^{-30}{\rm e\cdot cm},\nonumber\\
	&&|d_n|<1.8\times10^{-26}{\rm e\cdot cm}.
\end{eqnarray}
It is evident that the experimental upper bounds on the electron EDM are quite stringent, thereby imposing tight constraints on any new sources of CPV. On the other hand, investigating the impact of NP effects on the electron EDM could contribute significantly to the understanding of the mechanisms underlying CPV.

Among the myriad extensions of the SM, supersymmetry stands out as one of the most popular candidates. The EWBG analysis within the MSSM has been thoroughly discussed in Refs.~\cite{Dine:1990fj,Cohen:1992zx,Huet:1995sh,Chang:2002ex,Lee:2004we,Konstandin:2005cd,Chung:2008aya,Chung:2009qs,Chung:2009cb,Cirigliano:2009yd,Chiang:2009fs,Morrissey:2012db,Kozaczuk:2012xv}, while the EWBG analysis in non-minimal supersymmetric models has been elaborately addressed in Refs.~\cite{Pietroni:1992in,Davies:1996qn,Huber:2000mg,Kang:2004pp,Huber:2006wf}. In summary, the results indicate that the primary contributions to $Y_B$ predominantly arise from the T-term (the trilinear scalar term in the soft supersymmetry-breaking potential) and the $\mu$-term (the bilinear Higgs mass terms in the superpotential). The impact of supersymmetry effects on the electron EDM has been investigated in the referenced studies~\cite{Nath:1991dn,Kizukuri:1992nj,Falk:1996ni,Falk:1998pu,Brhlik:1998zn,Bartl:1999bc,Abel:2001vy,Barger:2001nu,Olive:2005ru,Cirigliano:2006dg,YaserAyazi:2006zw,Engel:2013lsa,Chupp:2017rkp,Ibrahim1998,Carena1998}. The results suggest that the cancellation of contributions from different phases is the most plausible explanation for suppressing the electron EDM below the corresponding experimental upper limits. Moreover, this cancellation is very stringent, generally reaching the level of $1\%$. When it is assumed that only the CPV phases originate from the T-term and the $\mu$-term, their values are strictly constrained by the experimental upper bounds. Overall, studying the EWBG and the electron EDM in a combined way in NP models is of great interest, and will deepen our understanding of CPV.

The TNMSSM is an extension that amalgamates two expansions of the MSSM, namely the NMSSM and the TMSSM. This model alleviates the little hierarchy problem by introducing two scalar triplets and makes the $\mu$ term coincide with the soft SUSY breaking scale naturally by introducing a scarlar singlet~\cite{agashe2011improving,basak2012triplet,basak2013130,search,77,88,9,99,10,11}. Correspondingly, new CPV sources are introduced in this model, and these TNMSSM specific CPV sources affect baryon asymmetry $Y_B$ and the electron EDM $d_e$. Hence, we focus on explaing the impacts of CPV sources in the TNMSSM on $Y_B$ and $d_e$ in this work.

The organization of this paper is as follows. In Section 2, we provide a brief description of the TNMSSM by introducing the superpotential and the general soft-breaking terms. Then, in Section 3, we analyze the electroweak phase transition (EWPT), the $Y_B$, the electron EDM $d_e$, the neutron EDM $d_n$ and mercury EDM $d_{Hg}$ within the TNMSSM. In Section 4, we explore the impact of CPV on the $Y_B$, electron EDM $d_e$, neutron EDM $d_n$ and mercury EDM $d_{Hg}$ by varying different parameters. The conclusions are summarized in Section 5.

\section{The TNMSSM\label{sec2}}

Like any general softly broken supersymmetric theory, the Lagrangian of the TNMSSM is characterized by the superpotential, supersymmetric gauge interactions, and various soft-breaking couplings (soft masses and trilinear terms).\;\,In the TNMSSM, a gauge singlet superfield S is introduced, along with two triplet superfields $T$ and $\overline{T}$.\;\,The corresponding superpotential is given by Ref.~\cite{agashe2011improving}
\begin{eqnarray}	
&&W = S \left( \lambda H_u \cdot H_d + \lambda_T \text{tr}(\bar{T} T) \right) + \frac{\kappa}{3} S^3 + \chi_u H_u \cdot \bar{T} H_u + \chi_d H_d \cdot T H_d\nonumber\\
&&\;\;\;\;\;\;\;\;+ Y_u H_u \cdot Q \bar{u} - Y_d H_d \cdot Q \bar{d} - Y_e H_d \cdot L \bar{e},
\end{eqnarray}
where $\lambda$, $\lambda_T$, $\kappa$, $\chi_u$, and $\chi_d$ are dimensionless Yukawa couplings. And the triplet superfields with hypercharge $Y=\pm1$ are defined as follows:
\begin{eqnarray}
T&\equiv& T^a \sigma^a =
\begin{pmatrix}
T^+/\sqrt{2}&-T^{++}\\
T^0&-T^+/\sqrt{2}
\end{pmatrix},\\
\bar{T}&\equiv& \bar{T}^a\sigma^a =
\begin{pmatrix}	
\bar{T}^-/\sqrt{2}&-\bar{T}^{0}\\
\bar{T}^{--}&-\bar{T}^-/\sqrt{2}
\end{pmatrix},
\end{eqnarray}
where $\sigma^a$ are the usual $2\times2$ Pauli matrices, and the respective definitions of the products between two $SU(2)_L$ doublets and between two $SU(2)_L$ doublets with one $SU(2)_L$ triplet are given as
\begin{eqnarray}
	&&H_u \cdot H_d \;\;= H_u^+H_d^--H_u^0H_d^0, \\
	&&H_u \cdot \bar{T} H_u = \sqrt{2}H_u^+H_u^0\bar{T}^--\left(H_u^0\right)^2\bar{T}^0-\left(H_u^+\right)^2\bar{T}^{--}, \\
	&&H_d \cdot T H_d =\sqrt{2}H_d^-H_d^0T^+-\left(H_d^0\right)^2T^0-\left(H_d^-\right)^2T^{++}.
\end{eqnarray}

The soft terms in the Lagrangian include:
\begin{eqnarray}
	-\cal{L}_{\textnormal{soft}}
	&=& m_{H_u}^2|H_u|^2+m_{H_d}^2|H_d|^2+m_{S}^2|S|^2+m_{T}^2\textnormal{tr}(|T|^2)+ m_{\bar{T}}^2\textnormal{tr}(|\bar{T}|^2) \nonumber \\
	&& +m_{Q}^2\left|Q \right|^2 +m_{\bar{u}}^2\left|\bar{u}\right|^2 +m_{\bar{d}}^2\left|\bar{d} \right|^2 +m_{L}^2\left|L \right|^2 +m_{\bar{e}}^2\left|\bar{e} \right|^2 \nonumber \\
	&&+(T_uQ\cdot H_u\bar{u}-T_dQ\cdot H_d\bar{d}-T_eL\cdot H_d \bar{e}  \nonumber \\
	&&  +A S H_u\cdot H_d + A_T S \ \textnormal{tr}(T\bar{T}) + \frac{A_{\kappa}}{3}S^3+ A_u H_u\cdot\bar{T}H_u+A_d H_d\cdot T H_d+h.c.). \label{eq:new}
\end{eqnarray}

The $SU(2)_L\bigotimes U(1)_Y$ electroweak symmetry breaking occurs when the neutral parts of Higgs fields obtain the VEVs
\begin{eqnarray}
	&&H_d^0=\frac{1}{\sqrt2} (v_1 + \Re {H_d^0} +i \Im {H_d^0}),\quad\; H_u^0=\frac{1}{\sqrt2} (v_2 +\Re {H_u^0} +i \Im {H_u^0}), \nonumber\\
	&&T^0\,=\frac{1}{\sqrt2} (v_{T_1} + \Re {T^0} +i \Im {T^0}),\quad\; \;{\bar T}^0\,=\frac{1}{\sqrt2} (v_{T_2} +\Re {{\bar T}^0} +i \Im {{\bar T}^0}), \nonumber\\
	&&S\;\;=\frac{1}{\sqrt2} (v_S +\Re S +i \Im S).
\end{eqnarray}

For convenience, we define $v_T^2=v_{T_1}^2+v_{T_2}^2,\; v^2=v_1^2+v_2^2$ and $\tan\beta^{'}=\frac{v_{T_2}}{v_{T_1}}$ in analogy to the ratio of the MSSM VEVs ($\tan\beta=\frac{v_2}{v_1}$).

In the TNMSSM, the masses of the Z and W bosons are modified due to the presence of the triplet VEVs
\begin{eqnarray}
	&&M_Z^2=\frac{{g_1}^2+g_2^2}{4}u^2 \equiv \hat{g}^2u^2,  \\
	&&M_W^2=\frac{{g_2}^2}{4}(v_1^2+v_2^2+2v_{T_1}^2+2v_{T_2}^2),  \\
	&&u^2= v_1^2+v_2^2+2v_{T_1}^2+2v_{T_2}^2 \approx (246\;{\rm GeV})^2,
\end{eqnarray}
where $g_1$ and $g_2$ represent the gauge coupling constants of $U(1)_Y$ and $SU(2)_L$ respectively.

As mentioned in the introduction, the TNMSSM model addresses the $\mu$ problem(s) and the little hierarchy problem by incorporating additional singlet and triplet superfields. Regarding the $\mu$ problem(s), the vacuum expectation value $v_s$ of the singlet field $S$ will generate effective $\mu$ term for the Higgs doublet and triplet with
\begin{eqnarray}
\mu = \lambda v_s,  \ \ \
\mu_T = \lambda_T v_s.
\end{eqnarray}
we take $\mu$, $v_s$ as inputs in the following analy analysis and $\lambda$ is obtained by
 $\lambda=\frac{\mu}{v_s}$. For the little hierarchy problem, it can be clearly seen that the coupling of the triplet to the $up$-type Higgs $H_u$ introduces additional quartic couplings in the Higgs (tree-level) potential without mixing with the $down$-type Higgs $H_d$:
\begin{eqnarray}
	V_{\textnormal{Higgs}} \ni\;\sim \chi_u^2(H_u)^4.
\end{eqnarray}

The introduction of two triplets and a singlet, in addition to solving the $\mu$ problem and the little hierarchy problem, also enables some other interesting work, such as studies related to Higgs decay and the transition magnetic moments of Majorana neutrinos~\cite{Zhu2024,Zhang2024,Hu2024,Chen2024}. When scalar triplets are introduced, neutrinos can acquire Majorana masses through the type-II see-saw mechanism.

Compared to the MSSM, the TNMSSM superpotential introduces an additional physical CPV phase through the $Arg(\chi_u\chi_d\kappa\lambda_T^*(\lambda^*)^2)$.\;\;Notably, this will have an impact on the theoretical predictions for $Y_B$ and electron EDM $d_e$ within the TNMSSM.

\section{EWBG and electron EDM in the TNMSSM\label{sec3}}

In this section, we present the mechanism by which a strong first-order PT is achieved in the TNMSSM model, as well as the methodology for calculating the baryon asymmetry $Y_B$ , the electron EDM $d_e$, the neutron EDM $d_n$ and the mercury EDM $d_{Hg}$.

\subsection{Electroweak phase transion\label{sec3.1}}
The EWPT is a pivotal process in the early universe's transition from a high-temperature symmetric phase to a low-temperature broken phase. This transition involves the development of the Higgs field vacuum expectation value, leading to the spontaneous breaking of electroweak symmetry. When the universe cooled below the electroweak scale temperature, broken phase bubbles began to nucleate in the symmetric plasma and rapidly expand.\;CP asymmetry was generated through interactions with the bubble walls, leading to a baryon number density asymmetry in the symmetric phase ahead of the bubble wall.\;These baryon number density asymmetries would diffuse into the region ahead of the bubble wall and affect electroweak sphaleron transitions, producing more baryons than antibaryons. Eventually, as the bubbles expanded and merged, the entire universe transitioned into the broken phase, with the baryon number being preserved within the bubble walls due to the suppression of sphaleron transition rates.\;\,Successful EWBG requires a strong first-order EWPT, which typically necessitates new physics beyond the SM to achieve~\cite{Athron2023,Wagner2023,Morrissey2012,Trodden1999}.

\begin{figure}
	\setlength{\unitlength}{1mm}
	\centering
	\includegraphics[width=3in]{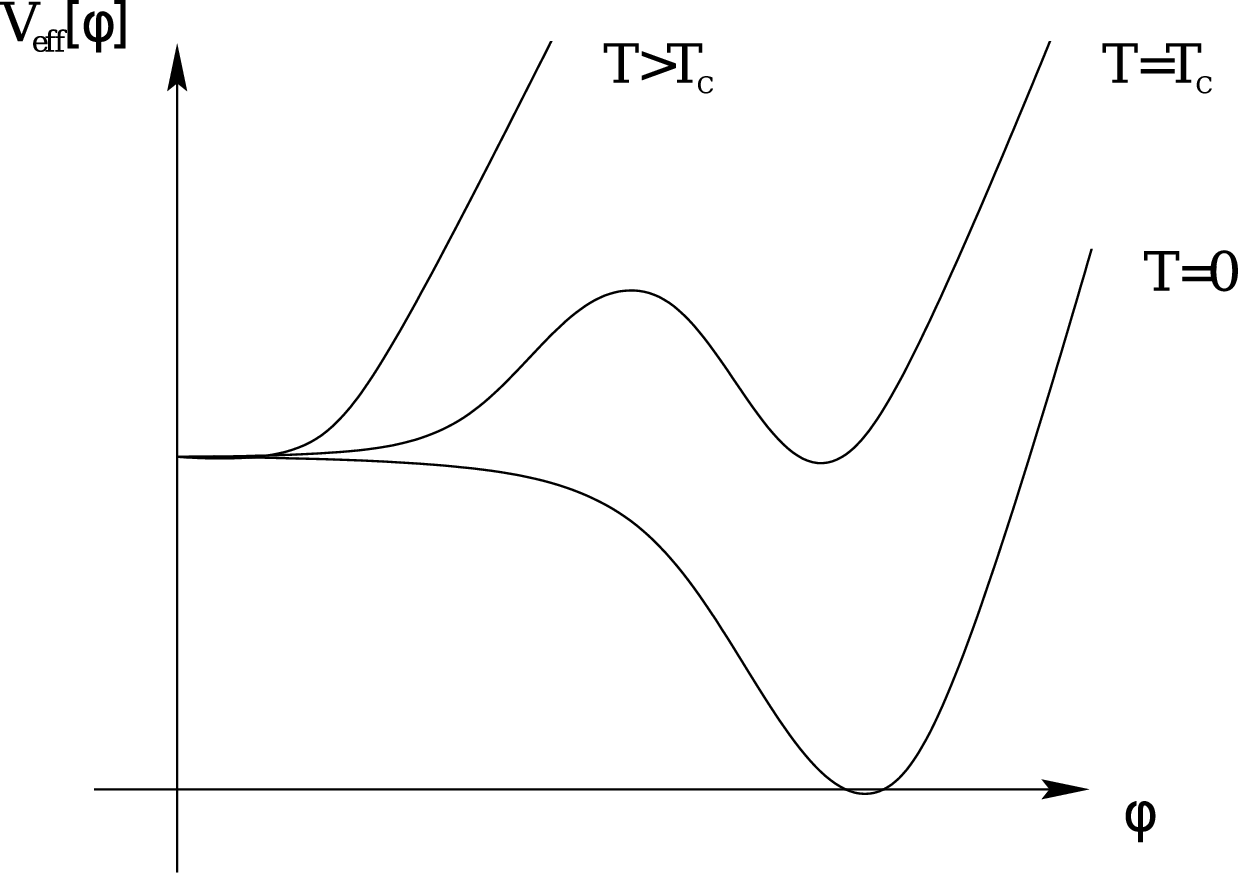}
	\vspace{0mm}
	\caption[]{First-order phase transition, the Higgs vacuum expectation value changes with the cooling of temperature as shown in the schematic diagram.\;\,The Higgs vacuum expectation value $v(T)$ at $T=T_C$ is defined as the temperature where the trivial and non-trivial minimum values are degenerate.
		From Ref.~\cite{Trodden1999}.}
	\label{fig:Firstorder}
\end{figure}

In the context of the MSSM, the discovery of the 125 GeV Higgs boson has rendered it improbable to have a strongly first-order EWPT with a very light right-handed stop of less than 120 GeV, thereby effectively ruling out the possibility of EWBG~\cite{Delepine:1996vn,Carena:1996wj,Espinosa:1996qw,Carena:1997ki,Huber:1998ck,Carena:2002ss,Profumo:2007wc,Carena:2008vj,Curtin:2012aa,Krizka:2012ah}. Compared to the MSSM, the TNMSSM, with the presence of the new singlet S, can realize a strong first-order phase transition.\;For simplicity, the temperature dependence of the Higgs VEVs $\beta$, $\beta'$ is neglected and the tree-level effective potential can be represented as
\begin{eqnarray}
	&&V_{eff}(h,\eta,S)=M(T)^2 h^2+m_{\eta}^2 \eta^2+m_s^2 S^2+\frac{1}{8}(g_1^2+g_2^2)(h^2cos2\beta+2\eta^2cos2\beta')^2\nonumber\\
	&&\qquad\qquad\;\;\;\;\;\,+\;\frac{2}{3}A_\kappa S^3,\label{Veff}
\end{eqnarray}
where
\begin{eqnarray}
&&M(T)^2\equiv M_0^2+\mathcal{Q}T^2\nonumber\\
&&\qquad\;\;\;\;\,=m_1^2\cos^2\beta+m_2\sin^2\beta+\lambda^2S^2+\kappa S^2\lambda \sin2\beta+\frac{1}{2}\lambda \sin2\beta\lambda_T\eta^2\sin^2\beta'\nonumber\\
&&\qquad\;\;\;\;\,+\;2\lambda_TS\eta \sin\beta'\chi_d\sin\beta
+2\lambda_TS\eta \cos\beta'\chi_u\cos^2\beta+4\chi_u^2\cos^2\beta\eta^2\sin^2\beta'\nonumber\\
&&\qquad\;\;\;\;\,+\;2\lambda S\sin2\beta\chi_u\eta \sin\beta'+2\lambda S\sin2\beta\chi_d\eta \cos\beta'-AS\sin2\beta-2A_u\cos^2\beta\eta \sin\beta'\nonumber\\
&&\qquad\;\;\;\;\,-\;2A_d\sin^2\beta\eta cos\beta'+4\chi_d^2\sin^2\beta \cos^2\beta'+\mathcal{Q}T^2,\label{MT}\\
&&\,m_\eta^2=\frac{1}{2}\lambda h^2\sin2\beta\lambda_T\sin^2\beta'-\kappa S^2\lambda_T \sin^2\beta'+\lambda_T^2S^2+4\chi_u^2\cos^2\beta\eta^2\sin^2\beta'\nonumber\\
&&\;\;\;\;\;\;+S\;4\chi_d^2\sin^2\beta \cos^2\beta'-A_tS\sin2\beta'+m_Tcos^2\beta'+m_{\bar T}^2\sin^2\beta',\\
&&\,m_S^2=\kappa\lambda h^2\sin2\beta-\kappa\lambda_T \sin^2\beta'+\lambda_T\eta^2+\lambda^2h^2+m_s^2,
\end{eqnarray}
and $\mathcal{Q}$ represents the sum of relevant thermal coupling, $T$ denotes temperature, $h$ and $\eta$ acquire VEVs $<h>=v$, $<\eta>=v_T$ respectively at zero temperature (present universe). At extremely high temperatures, $h$, $\eta$ and $S$ are stable at the origin. As the universe cooled, upon reaching the critical temperature for the phase transition, the Higgs field spontaneously broke the symmetry acquiring a non-zero vacuum expectation value, as shown in Fig.~\ref{fig:Firstorder}.

Furthermore, it can be noted in the Eqs. (\ref{Veff},$\;$\ref{MT}) indicate that the only gauge-independent term possible is $\mathcal{Q}T^2$, and the gauge independence of the $\mathcal{O}(T^2)$ term has been demonstrated in Appendix C of Ref.~\cite{Patel:2011th}. Therefore, the EWPT in our analysis is gauge invariant. In the early formation of the universe, as the temperature decreased, the Higgs field expectation value abruptly jumped from zero to a non-zero value, achieving a strong first-order PT at temperature $T\sim m_W$. Subsequently, we can obtain $T$ by solving the equations
\begin{eqnarray}
	&&\left\{\begin{array}{l}
	\ \frac{\partial V_{eff}}{\partial h}{\Big|}^{T_c}_{(v_c,v_{T_c})}=0,\\
		\\
		\mathcal\
		V_{eff}(h,\eta,S)=0.
	\end{array}\right.\;
\end{eqnarray}
As stated at the beginning of this section, for EWBG to occur, the sphaleron process must decouple once the phase transition is complete. In other words, at that moment, the sphaleron rate in the broken phase should be smaller than the Hubble parameter.\;In general, the sphaleron decoupling condition is cast into the form $v(T_C)/T_{C}\gtrsim1$.\;And in the parameter space selected in the following section of our analysis, we have $v(T_C)/T_{C}\gtrsim1.3$.\;This is sufficient for EWBG to occur.

\subsection{Baryon asymmetry $Y_B$\label{secA}}
Investigating the relationship between CPV effects and quantum transport equations is a complex and important aspect of research into EWBG. CPV effects act as source terms in the quantum transport equations that govern the generation of chiral charge at phase boundaries. In the TNMSSM model, the natural encoding of many CPV phases, such as $\kappa$, $\chi_d$, $\chi_d$, $\mu$, $T_u$, $T_d$, $T_e$, is considered. Among these, the phase $\mu$, $T_u$, $T_d$, $T_e$ have significant impact on the generation of $Y_B$. According to Ref.~\cite{Lee:2004we}, the expression for the baryon-to-entropy ratio can be simplified to
\begin{eqnarray}
	&&Y_B=F_1\sin \theta_\mu+F_2\sin (\theta_\mu+\theta_T)\label{F}.
\end{eqnarray}
where we have taken the gaugino mass terms $M_{1,2}$ to be real. And the phases $\mu$ and T are two independent CPV phases that influence baryon asymmetry and electron EDM $d_e$ within the model. The coefficients $F_i$ depend on the mass parameters in the TNMSSM, such as $\mu$, $M_1$, $M_2$, $T_0$ (we assume that T-terms are all same at the GUT scale, $T_u/Y_u=T_d/Y_d=T_e/Y_e=T_0$, where $Y_{u,d,e}$ are the corresponding Yukawa coupling constants) and squark masses $M_{\tilde t_L}$, $M_{\tilde t_R}$~\cite{Yang2020}. Furthermore, $F_i$ also have a overall dependence on bubble wall parameters $v_w$, $L_w$, $\Delta\beta$. For Eq. (\ref{F}), the detailed parameters used, such as the bubble wall parameter $v_w$, $L_w$ and $\Delta\beta$ are calculated with Ref.~\cite{Riotto1998}, the thermal width with Ref.~\cite{Huet1996,Akula:2017yfr}, and the diffusion constant is selected with Refs.~\cite{Enqvist1998,Braaten1992}. The relevant thermal parameters we have chosen will also be provided in Appendix~\ref{parameters}.

\subsection{The EDM of electron $d_e$, $d_n$ and $d_{Hg}$\label{secC}}

The effective Lagrangian for the electron EDM can be written as
\begin{eqnarray}
	&&\mathcal{L}_{EDM}=-\frac{i}{2}d_e\bar l_e\sigma^{\mu\nu}\gamma_5 l_e F_{\mu\nu},\label{MEDM}
\end{eqnarray}
where $\sigma^{\mu\nu}=i[\gamma^\mu,\gamma^\nu]/2$, $F_{\mu\nu}$ is the electromagnetic field strength, and $l_e$ denotes the electron which is on-shell.

\begin{figure}
	\setlength{\unitlength}{1mm}
	\centering
	\includegraphics[width=5in]{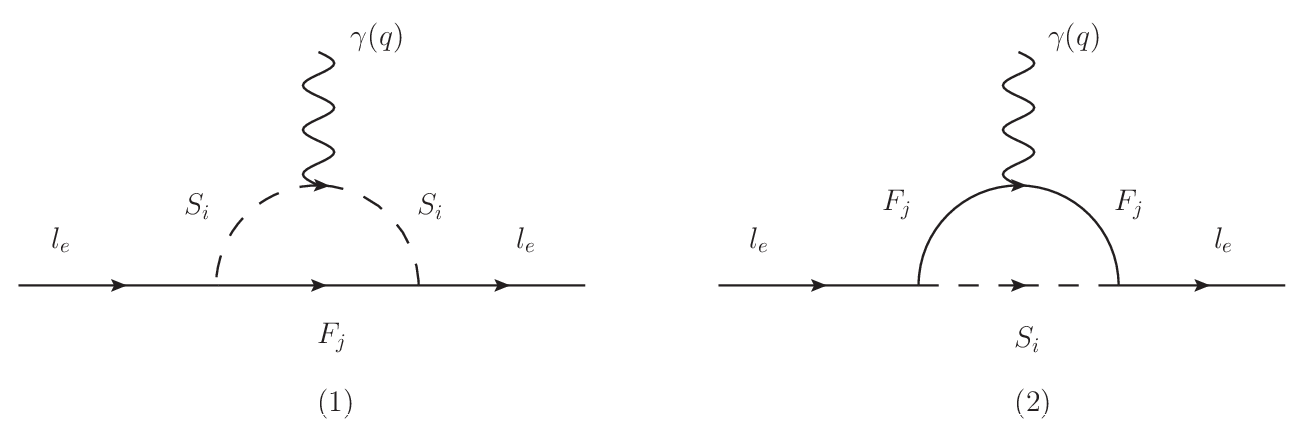}
	\vspace{0cm}
	\caption[]{The one-loop Feynman diagrams contributing to the electron EDM, where (1) denotes a charged scalar loop, and (2) denotes a charged fermion loop.}
	\label{Feynman diagram}
\end{figure}

When the internal mass $m_{_V}$ is much larger than the external electron mass $m_e$, the effective Lagrangian method can conveniently be used to obtain the contribution of loop diagrams to the EDM. Since ${/\!\!\! p}=m_e \ll m_{_V}$ for on-shell electron and ${/\!\!\! k}\rightarrow 0 \ll m_{_V}$ for photon, we can expand the amplitude of corresponding Feynman diagrams according to the external momenta of electron and photon. After matching the effective theory with the full theory, one obtains the high-dimensional operators and their coefficients. In subsequent calculations, it is sufficient to retain only these 6-dimensional operators~\cite{Feng,Feng1,Feng2,Zhang2014}:
\begin{eqnarray}
	&&O_2^{L,R} = {eQ_{_f}\over(4\pi)^2} \overline{(i \mathcal{D}_\mu l_e )} \gamma^\mu F\cdot \sigma P_{L,R} l_e, \nonumber\\
	&&O_3^{L,R} = {eQ_{_f}\over(4\pi)^2} \bar l_e  F\cdot \sigma \gamma^\mu P_{L,R} {(i \mathcal{D}_\mu l_e )}, \nonumber\\
	&&O_6^{L,R} = {eQ_{_f}m_e\over(4\pi)^2} \bar l_e F\cdot \sigma P_{L,R} l_e,
	\label{operators}
\end{eqnarray}
where $\mathcal{D}_\mu=\partial^\mu+ieA_\mu$, $P_L=\frac{1}{2}{(1 - {\gamma _5})}$, $P_R=\frac{1}{2}{(1 + {\gamma _5})}$, $e$ represents the electric charge, $Q_f=-1$ represents the charge number of electron and $m_e$ is the mass of electron $l_e$. The 6-dimensional operator in Eq.~(\ref{operators}) induce the effective coupling between the photon and the electron. The effective vertex with one external photon can be written as
\begin{eqnarray}
	&&O_2^{L,R} = {ieQ_{_f}\over(4\pi)^2} ({/\!\!\! p}+{/\!\!\! k})[{/\!\!\! k}, \gamma_\rho] P_{L,R}, \nonumber\\
	&&O_3^{L,R} = {ieQ_{_f}\over(4\pi)^2} [{/\!\!\! k}, \gamma_\rho] {/\!\!\! p} P_{L,R}, \nonumber\\
	&&O_6^{L,R} = {ieQ_{_f}m_e\over(4\pi)^2} [{/\!\!\!k}, \gamma_\rho]P_{L,R}.
\end{eqnarray}
If the full theory is invariant under the combined transformation of charge conjugation, parity and time reversal (CPT), the induced effective theory preserves the symmetry after the heavy freedoms are integrated out. The fact implies the Wilson coefficients of the operators $O_{2,3,6}^{L,R}$ satisfying the relations~\cite{Feng,Zhang2014}
\begin{eqnarray}
	C_3^{L,R}=C_2^{R,L\ast}, \qquad  C_6^{L}=C_6^{R\ast}\,,
\end{eqnarray}
where $C_i^{L,R}$ represent the Wilson coefficients of the corresponding operators $O_i^{L,R}$ in the effective Lagrangian. After applying the equations of motion to the external electron, we find that the concerned terms in the effective Lagrangian are transformed into
\begin{eqnarray}
	&&\quad\; C_2^{L}O_2^L +  C_2^{R\ast}O_3^L + C_6^{R\ast}O_6^L \nonumber\\
	&&\Rightarrow (C_2^{R\ast} + C_2^{L} + C_6^{R\ast})O_6^L \nonumber\\
	&&={ieQ_{_f}m_{_e}\over(4\pi)^2} \Im(C_2^{R} + C_2^{L\ast} + C_6^{R})\bar l_e \sigma^{\mu\nu} \gamma_5 l_e   F_{\mu\nu},
	\label{CtoRI}
\end{eqnarray}
where, $\Im(\cdots)$ denotes the operation to take the imaginary part of a complex number. Matching between Eq.~(\ref{MEDM}) and Eq.~(\ref{CtoRI}), we can obtain
\begin{eqnarray}
	&&d_e=-\frac{2eQ_fm_e}{(4\pi)^2}\Im(C_2^R+C_2^{L*}+C_6^R),
	\label{EDMe}
\end{eqnarray}
Next we will analyse the dominant one-loop diagrams contributing to the electron EDM $d_e$ in the TNMSSM, which are depicted by Fig.~\ref{Feynman diagram}. Calculating the Feynman diagrams, the electron EDM $d_e$ can be written as
\begin{eqnarray}
	&&d_e^{(1)}=\frac{-2}{em_e}\Im\Big\{x_e[-I_3(x_{F_j},x_{S_i})+I_4(x_{F_j},x_{S_i})][(C_{\bar l_e S_i F_j}^LC_{\bar F_j S_i l_e}^R)+(C_{\bar l_e S_i F_j}^RC_{\bar F_j S_i l_e}^L)^*]\nonumber\\
	&&\qquad\quad+\sqrt{x_{e}x_{F_j}}[-2I_1(x_{F_j},x_{S_i})+2I_3(x_{F_j},x_{S_i})]C_{\bar l_e S_i F_j}^RC_{\bar F_j S_i l_e}^R\Big\},\\
	&&d_e^{(2)}=\frac{-2}{em_e}\Im\Big\{x_e[-I_1(x_{F_j},x_{S_i})+2I_3(x_{F_j},x_{S_i})-I_4(x_{F_j},x_{S_i})][(C_{\bar l_e S_i F_j}^RC_{\bar F_j S_i l_e}^L)\nonumber\\
	&&\qquad\quad+(C_{\bar l_e S_i F_j}^LC_{\bar F_j S_i l_e}^R)^*]+\sqrt{x_{e}x_{F_j}}[2I_1(x_{F_j},x_{S_i})-2I_2(x_{F_j},x_{S_i})-2I_3(x_{F_j},x_{S_i})]\nonumber\\
	&&\qquad\quad\times C_{\bar l_e S_i F_j}^RC_{\bar F_j S_i l_e}^R\Big\},
\end{eqnarray}
where $x_i=m_i^2/m_W^2$, $C_{abc}^{L,R}$ denotes the constant parts of the interactional vertex about $abc$, which can be got through SARAH, the interacting particles are denoted by $a$, $b$, $c$. And the specific expressions for the functions $I_{1,2,3}(x_1,x_2)$ is given by Refs.~\cite{Zhang:2013hva,Zhang:2013jva}. For completeness, we will provide the specific forms of $C_{abc}^{L,R}$ and $I_{1,2,3}(x_1,x_2)$ in Appendix~\ref{C}.

\begin{figure}
	\setlength{\unitlength}{1mm}
	\centering
	\includegraphics[width=6in]{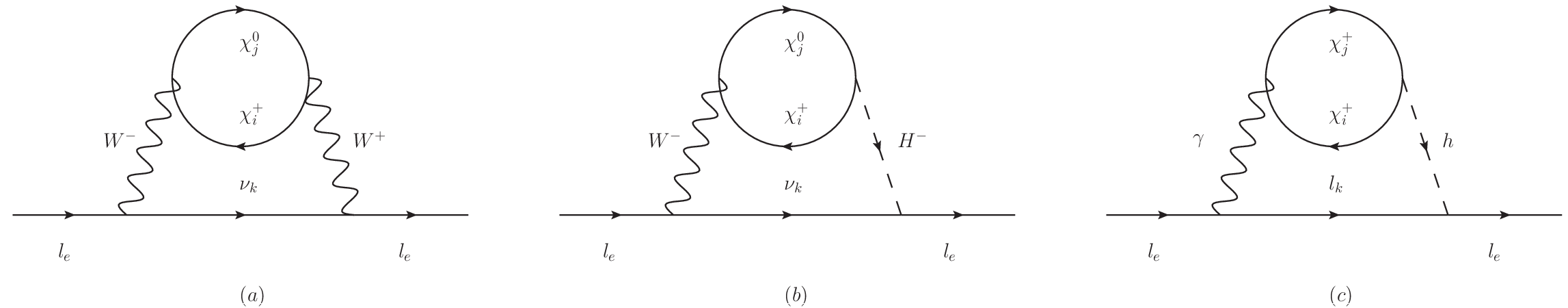}
	\vspace{0cm}
	\caption[]{The two-loop Barr-Zee diagrams, where one closed fermion loop is connected to virtual gauge bosons or the Higgs field, provide corresponding contributions to $d_e$ by attaching photons to the internal particles in all possible ways.}
	\label{Feynman diagram two-loop}
\end{figure}

The two-loop Barr-Zee diagrams can provide corrections to the electron EDM $d_e$. Here, we consider the contributions from two-loop diagrams where closed fermion loops are connected to virtual gauge bosons or the Higgs field. According to Ref.~\cite{Yang:2009zzh}, the Feynman diagrams are shown in Fig.~\ref{Feynman diagram two-loop}.
Assuming $m_F=m_{\chi^+_i}=m_{\chi^0_j}\gg m_W$, $m_F=m_{\chi^+_i}\gg m_{h}$, the contributions from the two-loop diagrams to $d_e$ can be simplify as
\begin{eqnarray}
	&&d_e^{(a)}=\frac{3G_F m_W\sqrt{x_{e}}}{-64\sqrt{2}\pi^4}\Big\{\Im(C_{\bar f_jWf_i}^LC_{\bar f_jWf_i}^{R*})\Big\},\nonumber\\
	&&d_e^{(b)}=\frac{G_F e m_{W} C_{\bar l_e H \nu_k}^L}{256\pi^4 g_2\sqrt{x_{_F}}}\Big\{\Big[179/36+10/3J(x_{_F},x_{_W},x_{_H})\Big]\Im(C_{\bar f_iHf_j}^L
	C_{\bar f_jWf_i}^L+C_{\bar f_iHf_j}^RC_{\bar f_jW\chi_i^+}^R)\nonumber\\
	&&\qquad\quad+\Big[-1/9-2/3
	J(x_{_F},x_{_W},x_{_H})\Big]\Im(C_{\bar f_iHf_j}^LC_{\bar f_jWf_i}^R+C_{\bar f_iHf_j}^RC_{\bar f_jWf_i}^L)\nonumber\\
	&&\qquad\quad+\Big[-16/9-8/3
	J(x_{_F},x_{_W},x_{_H})\Big]\Im(C_{\bar f_iHf_j}^LC_{\bar f_jWf_i}^L-C_{\bar f_iHf_j}^RC_{\bar f_jW\chi_i^+}^R)\nonumber\\
	&&\qquad\quad+\Big[-2/9-4/3
	J(x_{_F},x_{_W},x_{_H})\Big]\Im(C_{\bar f_iHf_j}^LC_{\bar f_jWf_i}^R-C_{\bar f_iHf_j}^RC_{\bar f_jWf_i}^L)\Big\},\nonumber\\
	&&d_e^{(c)}=\frac{G_F e m_{W} C_{\bar l_e h^0 l_e}}{64\pi^4 \sqrt{x_{_F}}}\Im(C_{\bar f_i h^0 f_i}^L)\Big[1+\ln\frac{x_{_F}}{x_{_h}}\Big],
\end{eqnarray}
where $f_j$, $f_i$ denote $\chi_j^0$ and $\chi_i^\pm$ respectively, $W$ denotes $W$ boson, $H$ denotes charged Higgs boson, $h$ denotes SM-like Higgs boson, the concrete expressions for the function $J$ can be found in Ref.~\cite{Yang:2018guw}, and
we will also provide it in the Appendix~\ref{C}.

In addition to the constraints from the electron EDM $d_e$ on the new model, the neutron EDM $d_n$ and the mercury EDM $d_{Hg}$ also impose strict limits on the parameter space. We briefly discuss the neutron EDM $d_n$ and the mercury EDM $d_{Hg}$ here.

The EDM of neutron $d_n$ can be expressed in terms of fundamental dipoles~\cite{Pospelov:2000bw}
\begin{eqnarray}
	&&d_n=(1\pm0.5)[1.4(d_d^\gamma-0.25d_u^\gamma)+1.1e(d_d^g+0.5d_u^g)]\pm(22\pm10){\;\rm MeV}C_5,
	\label{EDMdn}
\end{eqnarray}
where $d_q^\gamma$, $d_q^g$, $C_5$ denote the quark EDM of $q$ from the electroweak interaction, the CEDM of $q$ and the coefficient of Weinberg operator at the chirality scale respectively. Using the running from Refs.~\cite{Braaten:1990gq,Degrassi:2005zd} at one-loop, $d_{d,u}^{\gamma,g}$ and $C_5$ can be expressed in terms of $d_c^g$ at the scale $m_c$~\cite{Sala:2013osa}. The values of the coefficients $1\pm0.5$ and $22\pm10{\;\rm MeV}$ are respectively adopted as $0.5$ and $12{\;\rm MeV}$.

The effective Lagrangian for the quark EDMs can be written as
\begin{eqnarray}
	&&\mathcal{L}_{EDM}=-\frac{i}{2}d_q^\gamma\bar q \sigma^{\mu\nu}\gamma_5 q F_{\mu\nu},
\end{eqnarray}
where $\sigma^{\mu\nu}=i[\gamma^\mu,\gamma^\nu]/2$, $q$ is the wave function for quark, and $F_{\mu\nu}$ is the electromagnetic field strength. Adopting the effective Lagrangian approach, the quark EDMs can be written as
\begin{eqnarray}
	&&d_q^\gamma=-\frac{2e Q_q m_q}{(4\pi)^2}\Im(C_2^R+C_2^{L*}+C_6^R),
	\label{EDM}
\end{eqnarray}
where $C_{2,6}^{L,R}$ represent the Wilson coefficients of the corresponding operators $O_{2,6}^{L,R}$
\begin{eqnarray}
	&&O_2^{L,R}=\frac{e Q_q}{(4\pi)^2}(-iD_\alpha^*) \bar q \gamma^\alpha F\cdot\sigma P_{L,R}q,\nonumber\\
	&&O_6^{L,R}=\frac{e Q_q m_q}{(4\pi)^2}\bar q F\cdot\sigma P_{L,R}q.
\end{eqnarray}

\begin{figure}
	\setlength{\unitlength}{1mm}
	\centering
	\includegraphics[width=6.4in]{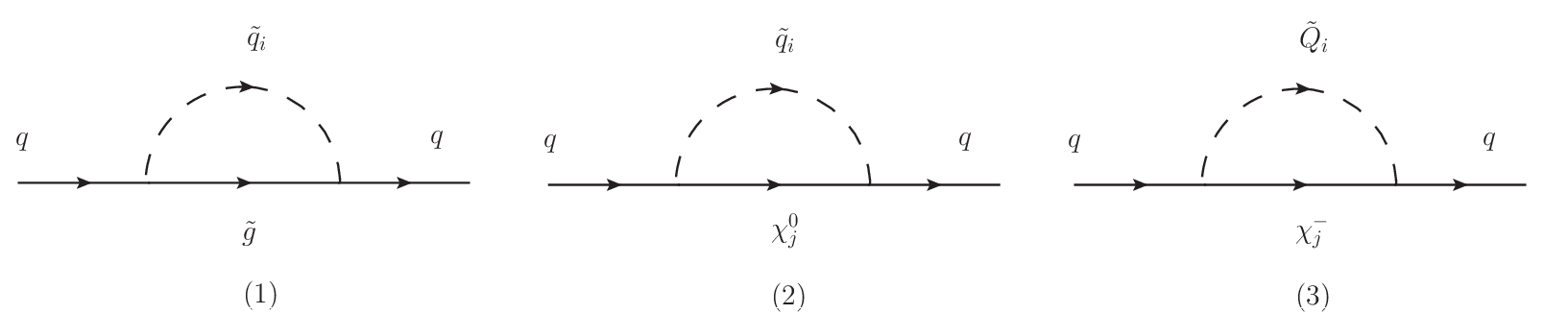}
	\vspace{0cm}
	\caption[]{The one-loop diagrams which contributes to $d_q^\gamma$ and $d_q^g$ are obtained by attaching a photon and a gluon respectively to the internal particles in all possible ways.}
	\label{Feynman diagram Q}
\end{figure}

Similarly, the effective Lagrangian for the quark CEDMs can be written as
\begin{eqnarray}
	&&\mathcal{L}_{CEDM}=-\frac{i}{2} d_q^g \bar q\sigma^{\mu\nu}\gamma_5 q G_{\mu\nu}^a T^a,
\end{eqnarray}
where $G_{\mu\nu}$ is the $SU(3)$ gauge field strength, $T^a$ is the $SU(3)$ generators. Then the quark CEDMs can be written as
\begin{eqnarray}
	&&d_q^g=-\frac{2g_3m_q}{(4\pi)^2}\Im(C_7^R+C_7^{L*}+C_8^R),
	\label{CEDM}
\end{eqnarray}
where $C_{7,8}^{L,R}$ represent the Wilson coefficients of the corresponding operators $O_{7,8}^{L,R}$
\begin{eqnarray}
	&&O_7^{L,R}=\frac{g_3}{(4\pi)^2}(-iD_\alpha^*) \bar q \gamma^\alpha G^a\cdot\sigma T^a P_{L,R}q,\nonumber\\
	&&O_8^{L,R}=\frac{g_3m_q}{(4\pi)^2}\bar q G^a\cdot\sigma T^a P_{L,R}q.
\end{eqnarray}
The one-loop Feynman diagrams contributing to the above amplitude are shown in Fig.~\ref{Feynman diagram Q}.
When calculating the Feynman diagrams, $d_q^\gamma$ and $d_q^g$ at the one-loop level can be written as
\begin{eqnarray}
	&&d_q^{\gamma(1)}=\frac{e_q e}{12\pi^2m_W}\frac{\sqrt {x_{\tilde g}}}{x_{\tilde q_i}}\Im\Big[C_{\bar{\tilde g} \tilde q_i q}^L C_{\bar q \tilde q_i \tilde g}^L\Big]I_1\Big(\frac{x_{\tilde g}}{x_{\tilde q_i}}\Big),\nonumber\\
	&&d_q^{g(1)}=\frac{-g_3}{32\pi^2m_W}\frac{\sqrt {x_{\tilde g}}}{x_{\tilde q_i}}\Im\Big[C_{\bar {\tilde g} \tilde q_i q}^L C_{\bar q \tilde q_i \tilde g}^L\Big]I_2\Big(\frac{x_{\tilde g}}{x_{\tilde q_i}}\Big),\nonumber\\
	&&d_q^{\gamma(2)}=\frac{e_q e}{32\pi^2m_W}\frac{\sqrt {x_{\chi_j^0}}}{x_{\tilde q_i}}\Im\Big[C_{\bar q \tilde q_i\chi_j^0}^L C_{\bar{\chi}_j^0 \tilde q_i q}^R\Big]I_1\Big(\frac{x_{\chi_j^0}}{x_{\tilde q_i}}\Big),\nonumber\\
	&&d_q^{g(2)}=\frac{g_3^3}{128\pi^2 e^2 m_W}\frac{\sqrt {x_{\chi_j^0}}}{x_{\tilde q_i}}\Im\Big[C_{\bar q \tilde q_i\chi_j^0}^L C_{\bar{\chi}_j^0 \tilde q_i q}^R\Big]I_1\Big(\frac{x_{\chi_j^0}}{x_{\tilde q_i}}\Big),\nonumber\\
	&&d_q^{\gamma(3)}=\frac{e}{16\pi^2m_W}\frac{\sqrt {x_{\chi_j^-}}}{x_{\tilde Q_i}}\Im\Big[C_{\bar q \tilde Q_i\chi_j^-}^L C_{\bar{\chi}_j^- \tilde Q_i q}^R\Big]\Big[e_QI_1\Big(\frac{x_{\chi_j^-}}{x_{\tilde Q_i}}\Big)+(e_q-e_Q)I_3\Big(\frac{x_{\chi_j^-}}{x_{\tilde q_i}}\Big)\Big],\nonumber\\
	&&d_q^{g(3)}=\frac{g_3^3}{16\pi^2e^2m_W}\frac{\sqrt {x_{\chi_j^-}}}{x_{\tilde Q_i}}\Im\Big[C_{\bar q \tilde Q_i\chi_j^-}^L C_{\bar{\chi}_j^- \tilde Q_i q}^R\Big]I_1\Big(\frac{x_{\chi_j^-}}{x_{\tilde Q_i}}\Big),
	\label{39}
\end{eqnarray}
where $x_i$ denotes $m_i^2/m_W^2$, $g_3$ is the strong coupling constant, $C_{abc}^{L,R}$ denotes the constant parts of the interaction vertex about $abc$ which can be obtained through SARAH, and $a$, $b$, $c$ denote the interacting particles. We will provide the specific forms of $C_{abc}^{L,R}$ and $I_{1,2,3}(x)$ in Appendix~\ref{C}.

\begin{figure}
	\setlength{\unitlength}{1mm}
	\centering
	\includegraphics[width=6in]{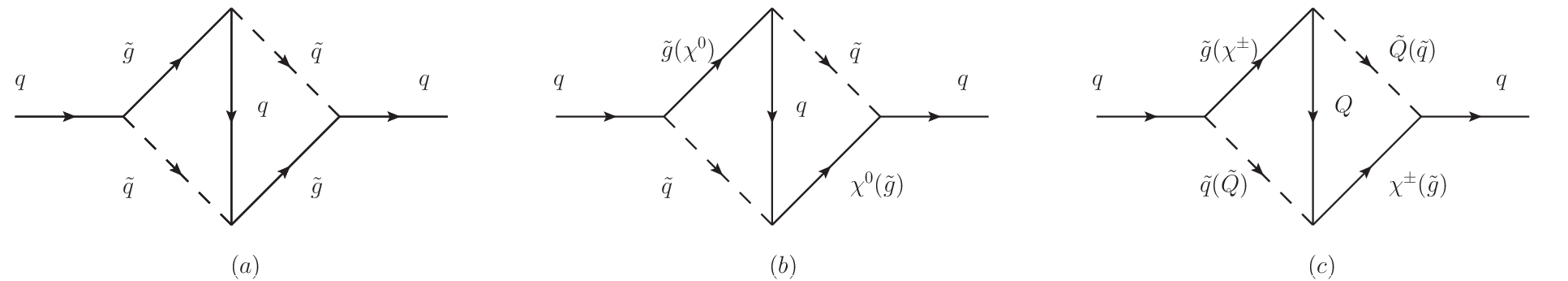}
	\vspace{0cm}
	\caption[]{The two-loop diagrams which contributes to $d_q^\gamma$ and $d_q^g$ are obtained by attaching a photon and a gluon respectively to the internal particles in all possible ways.}
	\label{two-loop Feynman diagram Q}
\end{figure}

The two-loop gluon corrections to the Wilson coefficients in the quark self-energy diagrams are considered, and the corresponding Feynman diagrams are shown in Fig.~\ref{two-loop Feynman diagram Q}. By attaching photons or gluons to internal particles in all possible ways, the corresponding dipole moment diagrams can be obtained.
The contributions of the two-loop diagrams to $d_q^\gamma$ and $d_q^g$ can be written as
\begin{eqnarray}
	&&d_q^{\gamma(a)}=\frac{-4e_q e g_3^2|m_{\tilde g}|}{9(4\pi)^4m_W^2}F_3(x_{_q},x_{_{\tilde q_j}},x_{_{\tilde g}},x_{_{\tilde g}},x_{_{\tilde q_i}})\Im[C_{\bar q \tilde g \tilde q_j}^L C_{\bar {\tilde g}q\tilde q_j}^L],\nonumber\\
	&&d_q^{g(a)}=d_q^{\gamma(a)}g_3/(e_q e),\nonumber\\
	&&d_q^{\gamma(b)}=\frac{4 e_q e}{3(4\pi)^4m_W^2}\Big\{|m_{\tilde g}|F_4(x_{_q},x_{_{\tilde q_j}},x_{_{\tilde g}},x_{_{\chi_k^0}},x_{_{\tilde q_i}})\Im[C_{\bar \chi_k^0 q \tilde q_j}^RC_{\bar \chi_k^0 q \tilde q_i}^L C_{\bar {\tilde g}q\tilde q_j}^LC_{\bar {\tilde g}q\tilde q_i}^L-C_{\bar \chi_k^0 q \tilde q_j}^L\nonumber\\
	&&\qquad\quad \times C_{\bar \chi_k^0 q \tilde q_i}^RC_{\bar q \tilde g \tilde q_j}^{L*}C_{\bar q \tilde g \tilde q_i}^{L*}]-m_{\chi_k^0}F_5(x_{_q},x_{_{\tilde q_j}},x_{_{\tilde g}},x_{_{\chi_k^0}},x_{_{\tilde q_i}})\Im[C_{\bar \chi_k^0 q \tilde q_j}^RC_{\bar \chi_k^0 q \tilde q_i}^RC_{\bar q \tilde g \tilde q_j}^{L*}C_{\bar {\tilde g}q\tilde q_i}^L\nonumber\\
	&&\qquad\quad-C_{\bar \chi_k^0 q \tilde q_j}^LC_{\bar \chi_k^0 q \tilde q_i}^LC_{\bar q \tilde g \tilde q_i}^{L*}C_{\bar {\tilde g}q\tilde q_j}^L]\Big\},\nonumber\\
	&&d_q^{g(b)}=d_q^{\gamma(b)}g_3/(e_q e),\nonumber\\
	&&d_q^{\gamma(c)}=\frac{2 e}{3(4\pi)^4m_W^2}\Big\{|m_{\tilde g}|F_4(x_{_Q},x_{_{\tilde Q_j}},x_{_{\tilde g}},x_{_{\chi_k^\pm}},x_{_{\tilde q_i}})\Im[C_{\bar Q \chi_k^\pm \tilde q_j}^LC_{\bar q\chi_k^\pm \tilde Q_i}^RC_{\bar{\tilde g}Q\tilde Q_j}^LC_{\bar{\tilde g}q\tilde q_i}^L-C_{\bar Q \chi_k^\pm \tilde q_j}^R\nonumber\\
	&&\qquad\quad\times C_{\bar q\chi_k^\pm \tilde Q_i}^LC_{\bar Q\tilde g\tilde Q_j}^{L*}C_{\bar q\tilde g\tilde q_i}^{L*}]-m_{\chi_k^\pm}F_5(x_{_Q},x_{_{\tilde Q_j}},x_{_{\tilde g}},x_{_{\chi_k^\pm}},x_{_{\tilde q_i}})\Im[C_{\bar Q \chi_k^\pm \tilde q_j}^LC_{\bar q\chi_k^\pm \tilde Q_i}^LC_{\bar Q\tilde g\tilde Q_j}^{L*}C_{\bar{\tilde g}q\tilde q_i}^L\nonumber\\
	&&\qquad\quad-C_{\bar Q \chi_k^\pm \tilde q_j}^RC_{\bar q\chi_k^\pm \tilde Q_i}^RC_{\bar{\tilde g}Q\tilde Q_j}^LC_{\bar q\tilde g\tilde q_i}^{L*}]\Big\},\nonumber\\
	&&d_q^{g(c)}=d_q^{\gamma(c)}g_3/e,
	\label{40}
\end{eqnarray}
where the concrete expressions for the functions $F_{3,4,5}$ can be found in Ref.~\cite{Feng:2004vu}. The similar expressions of two-loop gluino contributions can be found in Ref.~\cite{Feng:2004vu}. We write these symbols in a general form to facilitate the calculation of the quark EDM in other NP models.

\begin{figure}
	\setlength{\unitlength}{1mm}
	\centering
	\includegraphics[width=4in]{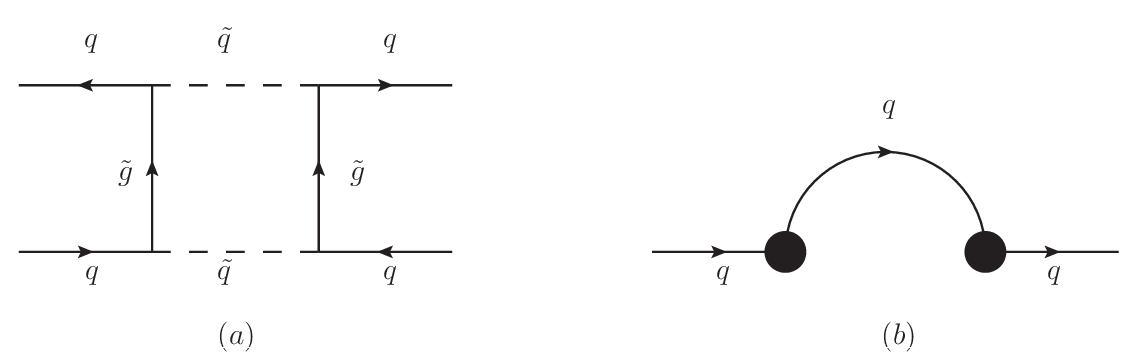}
	\vspace{0cm}
	\caption[]{Full theory diagram (a) and effective diagram (b) are plotted, where the blobs denote the effective vertexes, and a outgoing photon or gluon is attached by all possible ways.}
	\label{matching Feynman diagram}
\end{figure}

When SM quarks act as internal particles, there is an infrared divergence in Fig.~\ref{two-loop Feynman diagram Q}, because we calculate these diagrams by extending the external momenta. In this case, it is necessary to match the complete theory diagrams with the corresponding two-loop diagrams in Fig.~\ref{two-loop Feynman diagram Q} to eliminate the infrared divergence. Taking Fig.~\ref{two-loop Feynman diagram Q}(a) as an example, the method for eliminating infrared divergences is illustrated, with the corresponding schematic diagrams shown in Fig.~\ref{matching Feynman diagram}.
When an external gluon is attached to the internal particles in Fig.~\ref{two-loop Feynman diagram Q}(a), the external gluon can be attached to the same internal particles in Fig.~\ref{matching Feynman diagram}(a) or (b). Then, the infrared divergence in the diagram of Fig.~\ref{two-loop Feynman diagram Q}(a) with the attached gluon can be canceled by subtracting the corresponding diagrams in Fig.~\ref{matching Feynman diagram} with the gluon attached in the same manner.

\begin{figure}
	\setlength{\unitlength}{1mm}
	\centering
	\includegraphics[width=6in]{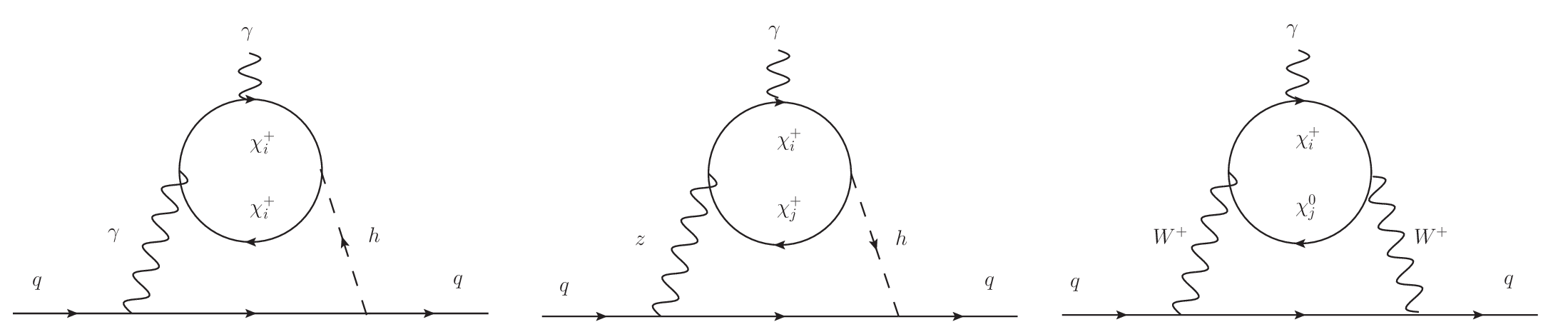}
	\vspace{0cm}
	\caption[]{The two-loop Barr-Zee type diagrams that contribute to the quark EDM. Diagrams where photons or gluons are emitted from the $W$ boson or the internal fermion do not contribute to the quark EDM or CEDM.}
	\label{two-loop Feynman Bazz}
\end{figure}

The two-loop Barr-Zee-type diagrams can also provide corrections to the quark EDM. Consider the diagrams where a closed fermion loop is connected to a virtual gauge boson or Higgs field, as shown in Fig.~\ref{two-loop Feynman Bazz}.
Then, the contribution of the two-loop Barr-Zee type diagrams to is given by~\cite{Giudice:2005rz}
\begin{eqnarray}
	&&d_q^{\gamma h}=\frac{e_q e^3}{32\pi^4 m_W}\frac{\sqrt{x_{\chi_i^+}}}{x_{h_k}}\Im\Big[C_{\bar\chi_i^+ h \chi_i^+}^R C_{\bar q h_k q}\Big]f_{\gamma H}\Big(\frac{x_{\chi_i^+}}{x_{h_k}}\Big),\nonumber\\
	&&d_q^{Zh}=\frac{e^2(T_{3q}-2e_q s_w^2)}{128\pi^4 c_w s_w m_W}\frac{\sqrt{x_{\chi_i^+}}}{x_{h_k}}\Im\Big[\Big(C_{\bar\chi_j^+ h_k \chi_i^+}^RC_{\bar\chi_i^+ Z \chi_j^+}^L-C_{\bar\chi_j^+ h_k \chi_i^+}^LC_{\bar\chi_i^+ Z \chi_j^+}^R\Big)C_{\bar q h_k q}\Big]\nonumber\\
	&&\qquad\;\;\;*f_{ZH}\Big(\frac{x_Z}{x_{h_k}},\frac{x_{\chi_i^+}}{x_{h_k}},\frac{x_{\chi_j^+}}{x_{h_k}}\Big),\nonumber\\
	&&d_q^{WW}=\frac{T_{3q} e^3}{128\pi^4 s_w^2 m_W}\sqrt{x_q x_{\chi_i^+}x_{\chi_j^0}}\Im\Big[C_{\bar\chi_j^0 W^+_\mu \chi_i^+}^RC_{\bar\chi_j^0 W^+_\mu \chi_i^+}^{L*}\Big]f_{WW}\Big(x_{\chi_i^+},x_{\chi_j^0}\Big),
	\label{41}
\end{eqnarray}
where $s_w\equiv\sin \theta_W,\;c_w\equiv\cos \theta_W$, and $\theta_W$ is the Weinberg angle, $T_{3q}$ denotes the isospin of the corresponding quark, the functions $f_{\gamma H},\;f_{ZH},\;f_{WW}$ can be found in Ref.~\cite{Giudice:2005rz}.

The EDM of mercury $d_{Hg}$ is primarily contributed by the CEDM of quarks ($d_q^g$), which can be written as~\cite{Lee:2004we}
\begin{eqnarray}
	d_{Hg} = - \left(d_d^g - d_u^g - 0.012 d_s^g \right)
	\times 3.2 \cdot 10^{-2} e  \ .
\end{eqnarray}
$d_q^g$ can be obtained in Eq.(\ref{39}), Eq.(\ref{40}) and Eq.(\ref{41}).

\section{Numerical analyses\label{sec4}}

In this section, we present and analyze the numerical results for the baryon asymmetry $Y_B$ and the electron EDM $d_e$ in the TNMSSM. The appropriate SM input parameters are selected as $m_W=80.3692\;{\rm GeV},\;m_Z=91.1880\;{\rm GeV},\;\alpha_{em}(m_Z)=1/137,\;\alpha_s(m_Z)=0.118$, we take the stop mass $m_{\tilde t_L}=m_{\tilde t_R}=1.5\;{\rm TeV}$, charged Higgs boson mass $M_{H^\pm}=1.5\;{\rm TeV}$ for simplicity. According to Refs.~\cite{AlvarezGaume:1983gj,Strumia:1996pr,Profumo:2006yu}, the magnitude of the scalar trilinear coupling is constrained by avoiding charge and color breaking minima.\;Thus, we take $T_{0}=0.1\;{\rm TeV}$ to satisfy these conditions. The mass of the SM-like Higgs boson is $125 \pm0.2\;{\rm GeV}$~\cite{ParticleDataGroup2024}, which imposes stringent constraints on the parameter space.\;Compared to the MSSM, the new triplet and singlet particles mix with the MSSM scalar doublets at the tree level in the TNMSSM, thereby affecting the theoretical predictions for the mass of the SM-like Higgs boson. And within our chosen parameter space, the lightest Higgs mass in the TNMSSM is constrained in the range
$124\;{\rm GeV}<m_h<126\;{\rm GeV}$. Within our parameter space, we choose $\tan\beta=2$, $\tan\beta'=0.8$, $M_{\tilde L,\tilde e}={\rm diag}(2,2,2)\;{\rm TeV}$. Taking into account the experimental constraints~\cite{ParticleDataGroup2024}, we appropriately fix
$M_1=600\;{\rm GeV}$ and $M_2=800\;{\rm GeV}$ for simplicity.

Considering the tree-level masses of $W$ boson and $Z$ boson, in the TNMSSM, the parameter $\rho$ can be written as~\cite{Yang:2024uoq}
\begin{eqnarray}
	\rho =\frac{m_{W}^2}{\cos^2\theta_W m_{Z}^2}=1-\frac{2v_T^2}{v^2},
\end{eqnarray}
where $\theta_{W}$ is the Weinberg angle. The constraint on this parameter within $2\sigma$-deviation requires $\rho_{\rm{exp}}=1.00038\pm0.00020\times2$ experimentally. Therefore, the scenario in which the value of $v_T^2$ does not exceed $1~\rm{GeV}$ is safer.

\begin{figure}
	\setlength{\unitlength}{1mm}
	\centering
	\includegraphics[width=3.1in]{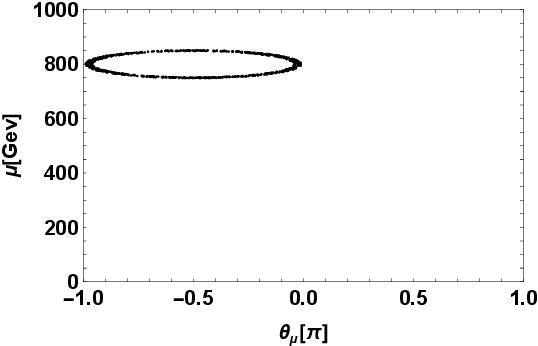}%
	\vspace{0.3cm}
	\caption[]{Keeping $Y_B$ in the region $(8.2-9.4)\times10^{-11}$, the allowed region of $\theta_\mu$ and $\mu$.}
	\label{13}
\end{figure}

In order to see how $\theta_\mu$, $\theta_{T_0}$ and $\mu$ affect $Y_B$, we scan the regions of the parameter space [$\theta_\mu=(-\pi,\pi),\;\theta_{T_0}=(-\pi,\pi),\;\mu=(100,1000)\;{\rm GeV}$]. During the scanning process, we keep $Y_B$ within the range of $(8.2-9.4)\times10^{-11}$. The allowed regions for $\theta_\mu$ and $\mu$ are shown in Fig.~\ref{13}. From the figure, it is evident that an ellipse is formed in the negative $\theta_\mu$ phase space near $\mu=800\;{\rm GeV}$. According to the Refs.~\cite{Lee:2004we,Riotto1998,Carena:1997gx}, a resonance behavior is induced when $\mu$ is approximately equal to $M_2$, which can lead to an enhancement of the $Y_B$ value. This is the reason why the ellipse is concentrated near $\mu=800\;{\rm GeV}$.
The concentration in the negative phase space arises because, in our Eq.(\ref{F}), the $F_1$ term contributed by the $S^{CP}_{ H}$ source is negative, while $Y_B$ is positive. This necessitates that the product $F_1\sin\theta_\mu$ be positive, which naturally leads to the phase angle $\theta_\mu$ being confined to the negative phase space. Through the memory effect, the required phase angle $\theta_\mu$ can be reduced by two orders of magnitude. An elaborate explanation of this memory effect is detailed in Ref.~\cite{Riotto1998}, and the expression for determining whether it can occur can be succinctly expressed as
\begin{eqnarray}
	&&|\sin \theta_\mu|\gtrsim 10^{-3} (\frac{v_w}{0.1}).
\end{eqnarray}
it represents that the observed baryon asymmetry may be caused by Higgsinos. This reasonably explains why a small phase can successfully generate $Y_B$.

\begin{figure}
	\setlength{\unitlength}{1mm}
	\centering
	\includegraphics[width=3.1in]{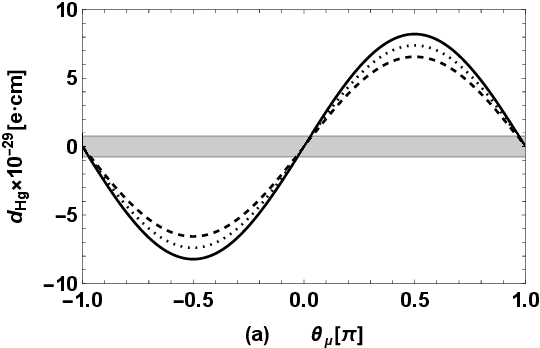}%
	\vspace{0.3cm}
	\includegraphics[width=3.1in]{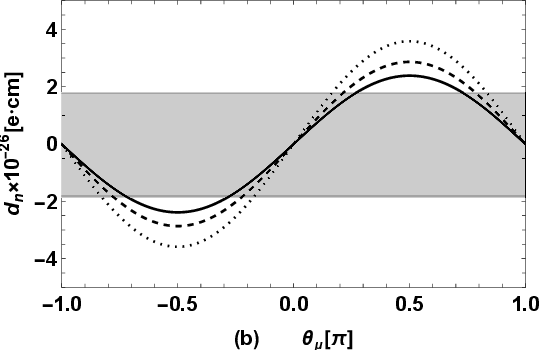}
	\vspace{0cm}
	\caption[]{(a) and (b) correspond to the results of the $d_{Hg}$ and $d_n$ scans in the parameter space, respectively. The solid line, dashed line, and dotted line correspond to $\mu=800\;{\rm GeV}$, $900\;{\rm GeV}$, $1000\;{\rm GeV}$, respectively. The gray bands represent the experimental upper limits.}
	\label{dnhg}
\end{figure}

To better explain the EWBG, we have scanned the neutron EDM $d_n$ and the mercury EDM $d_{Hg}$ in the parameter space. The results are shown in the Fig.~\ref{dnhg}.
This indicates that after considering the neutron EDM $d_n$ and the mercury EDM $d_{Hg}$, the phase space of parameter $\mu$ will be further compressed.

\begin{figure}
	\setlength{\unitlength}{1mm}
	\centering
	\includegraphics[width=3.1in]{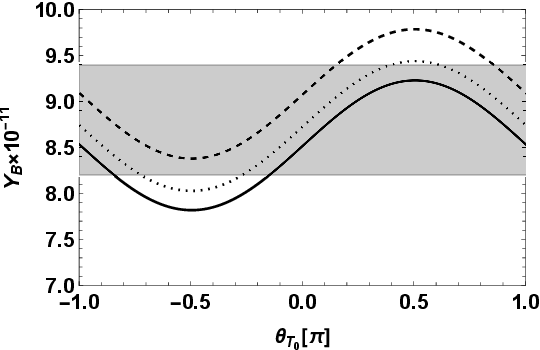}%
	\vspace{0.3cm}
	\caption[]{$Y_B$ versus $\theta_{T_0}$ for $\lambda_T=0.2$ (solid line), $0.3$ (dashed line), $0.4$ (dotted line), and the gray area denotes the experimental interval $(8.2-9.4)\times10^{-11}$.}
	\label{11}
\end{figure}

Then we select $\mu=800\;{\rm GeV}$ and $|\theta_\mu|=0.013$ to observe the effect of phase angle $\theta_{T_0}$ and new triplet couplings $\lambda_T$ on $Y_B$. $Y_B$ versus $\theta_{A_0}$ for $\lambda_T=0.2$ (solid line), $0.3$ (dashed line), $0.4$ (dotted line) is plotted in Fig.~\ref{11}, where the gray area denotes the experimental interval $(8.2-9.4)\times10^{-11}$. From the experimental results, the sensitivity of $Y_B$ to $\theta_{A_0}$ is less than its sensitivity to the new triplet coupling parameter $\lambda_T$. This indicates that while the contribution of the $S^{CP}_{t}$ source to $Y_B$ is not as significant as that of the $S^{CP}_{ H}$ source, it still exerts a certain influence on the outcome.

\begin{figure}
 	\setlength{\unitlength}{1mm}
 	\centering
 	\includegraphics[width=3.1in]{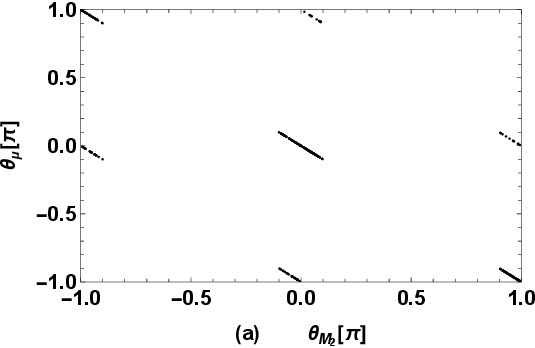}%
 	\vspace{0.3cm}
 	\includegraphics[width=3.1in]{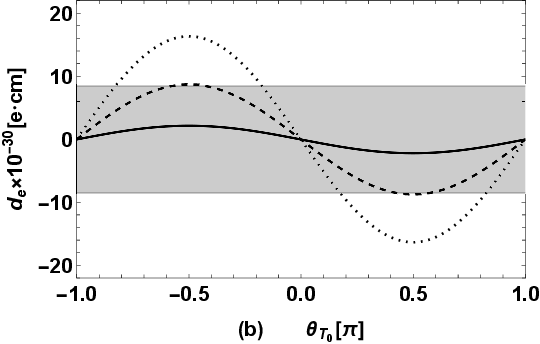}
 	\vspace{0cm}
 	\caption[]{Keeping $|d_e|<4.1\times10^{-30}$, the cancellation between $\theta_\mu$ and $\theta_{M_2}$ (a) are shown. And $d_e$ versus $\theta_{T_0}$ are plotted for $T_0=0.1\;{\rm TeV}$ (solid line), $0.3\;{\rm TeV}$ (dotted line), $0.5\;{\rm TeV}$ (dashed line), where the gray area denotes the present experimental upper bound on $d_e$.}
 	\label{theta-mu}
 \end{figure}

 To better illustrate the stringent constraints on the new CP source contributions within the model, we have conducted a discussion on the electron EDM $d_e$, similar to that in Refs.~\cite{Yang2020,Lee:2004we}.\;The numerical results for the $Y_B$ indicate that when EWBG occurs, we chose $|\theta{_\mu}|$ as $0.013$. However, in this scenario, the contribution to the electron EDM $d_e$, is significantly enhanced, with electron EDM $d_e$ exceeding the corresponding upper bound by several orders of magnitude.
 Therefore, to satisfy the current experimental upper limits, we require contributions from different CPV terms to cancel each other out. Subsequently, we set other CPV phases to zero to explore the cancellation between $\theta_\mu$ and $\theta_{M_2}$.\;We perform a scan over the parameter space [$\theta_\mu=(-\pi,\pi),\;\theta_{M_2}=(-\pi,\pi)$] while maintaining $|d_e|<4.1\times10^{-30}$. The numerical results are shown in Fig.~\ref{theta-mu} (a). Furthermore, $\theta_{T_0}$ can also make a significant contribution to the $Y_B$, hence it is interesting to explore how $\theta_{T_0}$ affects $d_e$. Due to the suppression of the impact of $\theta_{T_0}$ by the smaller $Y_e$, it is not necessary to counteract the contribution of $\theta_{T_0}$ to $d_e$. Subsequently, in  Fig.~\ref{theta-mu}(b), we plot the relationship between $d_e$ and $\theta_{T_0}$, where the gray area represents the experimental upper limit for $d_e$, and the solid, dotted, and dashed lines correspond to $T_0=0.1\;{\rm TeV},\;0.3\;{\rm TeV},\;0.5\;{\rm TeV}$, respectively.

 From the figure, it can be observed that when $\theta_\mu$ is confined to a small phase space, the contributions from $\theta_\mu$ and $\theta_{M_2}$ cancel each other out, which also indicates the contributions of $\theta_\mu$ and $\theta_{M_2}$ are comparable, since $\theta_\mu$ is the primary source of baryon asymmetry, concerns about whether the contribution to the electron EDM $d_e$ from the large $\theta_\mu$ values required for generating EWBG can be canceled can be relaxed. Furthermore, the contribution from $\theta_{T_0}$ is amplified by a large $T_0$, while the contributions from $\theta_\mu$ and $\theta_{M_2}$ are several orders of magnitude larger than that from $\theta_{T_0}$.\;Therefore, it is hard for $\theta_{T_0}$
 to cancel out the contribution from $\theta_\mu$
 (when $\theta_\mu$ and $\theta_{T_0}$ are canceled out, the maximum value of $\theta_\mu$ is ${\mathcal O}(10^{-3})$, which is insufficient to generate EWBG). This is in contrast to the situation in the MSSM~\cite{YaserAyazi:2006zw}, where when the contributions to the electron EDM $d_e$ from $\theta_\mu$
 and $\theta_{T_0}$ cancel each other out, the maximum value of $\theta_\mu$ can be large enough to induce EWBG. It can be seen that within our chosen parameter space, the contribution from the slepton is highly suppressed by the large slepton mass.

\begin{figure}
	\setlength{\unitlength}{1mm}
	\centering
	\includegraphics[width=3.1in]{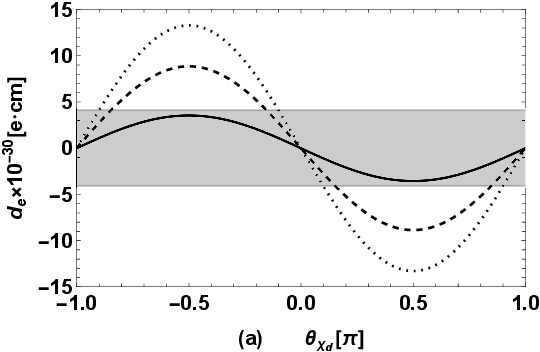}%
	\vspace{0.3cm}
	\includegraphics[width=3.1in]{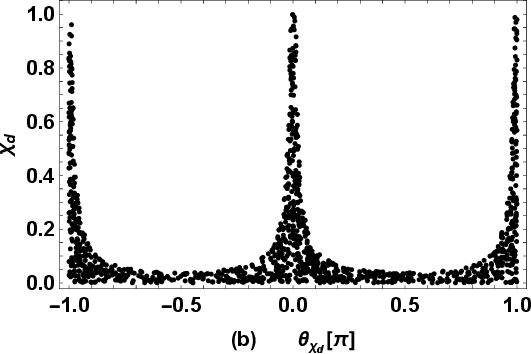}
	\vspace{0cm}
	\caption[]{$\chi_d$ effect on electron EDM $d_e$ contribution. (a) is the curve drawn at different times of input $\chi_d$, $0.01$ (solid line), $0.02$ (dashed line), $0.03$ (dotted line). The shadow band is the experimental limit. And (b) is the distribution of the scanned phase space $\chi_d$ from 0 to 1 when $|d_e|<4.1\times10^{-30}$ is maintained.}
	\label{d}
\end{figure}

\begin{figure}
	\setlength{\unitlength}{1mm}
	\centering
	\includegraphics[width=3.1in]{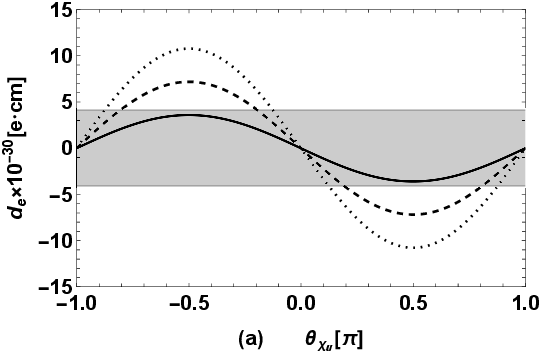}%
	\vspace{0.3cm}
	\includegraphics[width=3.1in]{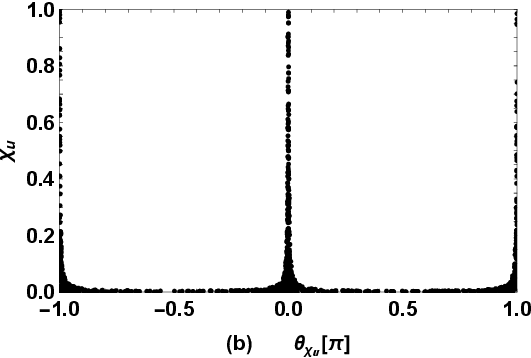}
	\vspace{0cm}
	\caption[]{$\chi_u$ effect on electron EDM $d_e$ contribution. (a) is the curve drawn at different times of input $\chi_u$, $0.001$ (solid line), $0.002$ (dashed line), $0.003 $ (dotted line). The shadow band is the experimental limit. And (b) is the distribution of the scanned phase space $\chi_u$ from 0 to 1 when $|d_e|<4.1\times10^{-30}$ is maintained.}
	\label{u}
\end{figure}

 In the TNMSSM, the three new types of CPV phases, $\kappa$, $\chi_d$ and $\chi_u$, also contribute to the electron EDM $d_e$. Among them, the contribution of $\kappa$ is constrained by the coupling to singlet with smaller degree of freedom, and is insufficient to affect the final result of the EDM. Hence we do not delve into a detailed discussion of it here. To better elucidate the influence of parameters $\chi_d$ and $\chi_u$ on the electron EDM $d_e$, we have conducted separate graphical analyses for each parameter, with the results presented in Fig.~\ref{d}, and Fig.~\ref{u}, respectively.

 Fig.~\ref{d} (a) depicts the parameter space scanned while holding all other CPV phases to zero and fixing $\chi_d$ at $0.01$ (solid line), $0.02$ (dashed line), $0.03$ (dotted line), respectively, with the phase angle $\theta_{\chi_d}$ varying from $-\pi$ to $\pi$. Fig.~\ref{d} (b) is keeping $|d_e|<4.1\times10^{-30}$ and scan the regions of the parameter space [$\theta_{\chi_d}=(-\pi,\pi),\;\chi_d=(0,1)$]. Fig.~\ref{u} (a) and (b) follow a similar approach, with the distinction that the parameter inputs related to $\chi_d$ have been replaced with those pertaining to $\chi_u$. Fig.~\ref{d} (b) and Fig.~\ref{u} (b) reveal that the points satisfying the experimental constraints for $\chi_d$ and $\chi_u$ form a narrow band at the base, with prominent peaks occurring near $\pm\pi$ and $0$. This indicates that small values of $\chi_d$ and $\chi_u$ can satisfy the experimental constraints across the entire CPV phase space, whereas larger values of $\chi_d$ and $\chi_u$ only meet these constraints when $\sin\theta_{\chi_d}$ and $\sin\theta_{\chi_u}$ approache zero. This is also reflected in  Fig.~\ref{d} (a) and  Fig.~\ref{u} (a). The solid lines, representing smaller values of $\chi_d$ and $\chi_u$, are observed to lie entirely within the shaded region, whereas the dashed and dotted lines, indicative of larger values of $\chi_d$ and $\chi_u$, only fall within the shaded region near $\pm\pi$ and $0$. Fig.~\ref{d} (b) and Fig.~\ref{u} (b) also reveal that when satisfying the experimental constraints, the band generated at the bottom of the phase space for $\chi_u$ is slightly lower than that for $\chi_d$. This shows that the upper limit of $\chi_u$ satisfying the experimental limit in the entire phase space is less than the upper limit of $\chi_d$. In other words, within the context of contributions to the electron EDM $d_e$, the parameter $\chi_u$ exhibits greater sensitivity compared to $\chi_d$. This is also reflected in the relevant parameter input in Fig.~\ref{d} (a) and Fig.~\ref{u} (a), where the parameter input for $\chi_u$ is an order of magnitude smaller than that for $\chi_d$.

\section{Summary\label{sec5}}

In this study, we focus on examining the impact of CPV sources on EWBG and the electron EDM $d_e$ within the TNMSSM. Compared to the MSSM, the new triplet and singlet states mix with the MSSM doublet states at the tree level, and under such circumstances, a strong one-step PT can be realized. When resonance enhancement occurs, generating EWBG requires $|\theta_\mu|\gtrsim0.013$, which makes significant contributions to the electron EDM. In this case, the cancellation effect is necessary to satisfy the experimental upper bound on the electron EDM. It is found that the contributions from $\theta_{M_2}$ ($M_2$ is the chargino mass term) to the electron EDM can cancel the ones from large $\theta_\mu$ required by generating EWBG. When $\theta_\mu$ and $\theta_{M_2}$ cancel each other out, $\theta_\mu$ will be confined to a small phase space by the expeirmental bound on the electron EDM. Besides the traditional CPV sources $\mu$ and $M_2$ in SUSY models, the new CPV sources $\chi_u$, $\chi_d$, and $\kappa$ can also make contributions to the electron EDM $d_e$, where $\chi_u$ and $\chi_d$ affect the theoretical prediction on $d_e$ acutely, while the effects of $\kappa$ is mild. This indicates that the cancellation phase between $\theta_\mu$ and $\theta_{M_2}$, which is constrained to a small phase space,  can be relaxed slightly by considering the effects of $\chi_u$ and $\chi_d$.

\begin{acknowledgments}
	
	The work has been supported by the National Natural Science Foundation of China (NNSFC) with Grants No. 12075074, No. 12235008, Hebei Natural Science Foundation with Grant No. A2022201017, No. A2023201041, the youth top-notch talent support program of the Hebei Province.

\end{acknowledgments}

\appendix

\section{A simple derivation of the baryon-to-entropy density ratio $Y_B$. \label{parameters}}

Here, we briefly introduce Eq. (\ref{F}) from Ref.~\cite{Lee:2004we}.

First, the baryon-to-entropy density ratio can be expressed as
\begin{eqnarray}
Y_B \equiv \frac{\rho_B}{s}\ ,
\end{eqnarray}
where $s$ represents the baryon entropy density at the time of electroweak phase transition
\begin{eqnarray}
s = (2 \pi^2)/45 \times  g_{\rm eff} (T)\ ,
\end{eqnarray}
$\rho_B$ is the baryon density
\begin{eqnarray}
\rho_B =\frac{n_F}{2} {\cal A}  \
\left[ r_1  \Gamma_{\rm ws}
+ r_2 \, \frac{\Gamma_{\rm ws}}{\Gamma_{\rm ss}}\frac{v_w^2}{\bar D}
\left(1- \frac{D_q}{\bar{D}} \right) \right] \,
\frac{\bar D}{v_w^2  + {\cal R}(\bar{D} + D_q)}\ ,\label{A3}
\end{eqnarray}
$n_F$ is the number of fermion families, $v_w$ is the wall velocity, $\bar{D}$ is an effective diffusion constant, $D_q$ is the quark diffusion constant. $\Gamma_{\rm ws}, \Gamma_{\rm ss}$ can be written as respectively
\begin{eqnarray}
&&\Gamma_{\rm ws} = 6 \kappa\alpha_w^5 T\ , \kappa \simeq 20\nonumber\ ,\\
&&\Gamma_{ss}=6 \kappa' \frac{8}{3} \alpha_{s}^4 T\ , \kappa' \sim \mathcal{O}(1)\ ,
\end{eqnarray}
$\alpha_w$ and $\alpha_s$ are the weak and strong fine-structure constants, respectively.

$r_1$ and $r_2$ are related combinations of the statistical factors $k_i$
\label{eq:rhob4b}
\begin{eqnarray}
r_1 &= \frac{9 k_Q k_T - 5 k_Q k_B - 8  k_T k_B}{k_H (9k_Q+9k_T+k_B)}\ , \nonumber\\
r_2 &=
\frac{k_B^2 (5k_Q+4k_T)(k_Q + 2 k_T)}{k_H (9k_Q+9k_T+k_B)^2} \ ,
\end{eqnarray}
with
\begin{eqnarray}
k_i(m_i/T) = k_i(0)\frac{c_{F,B}}{\pi^2}\int_{m/T}^\infty dx\,x\,
\frac{e^x}{(e^x \pm 1)^2}\sqrt{x^2 - m^2/T^2}\ ,
\end{eqnarray}
where for fermions (bosons) $c_{F(B)} = 6\,(3)$.
$\cal R$ is the relaxation term
\begin{eqnarray}
{\cal R} = \Gamma_{\rm ws}\, \left[
\frac{9}{4} \, \left(1 + \frac{ n_{\rm sq}}{6}\right)^{-1} + \frac{3}{2}
\right],
\end{eqnarray}
where $n_{\rm sq}$ indicates the number of light squark flavors.
In Eq. (\ref{A3}), $\cal A$ is a solution that satisfies the Higgs density equation. Higgs density equation can be written
\begin{eqnarray}
{\dot H} -{\bar D} \nabla^2 H +{\bar \Gamma} H-{\bar S}=
\mathcal{O}(\delta_{ss}, \delta_Y)\ ,\label{A7}
\end{eqnarray}
where $\bar{\Gamma}$ is an effective decay constant and $\bar S$ is an effective source term. Ref.~\cite{Lee:2004we} provides constraints on Eq. (\ref{A7}), which are
\begin{itemize}
	\item To simplify the problem to a one-dimensional issue, neglect the curvature of the wall, where all relevant functions depend on the variable $\bar{z} = |{x} + {v}_{w} t|$. Thus,  $\bar{z} < 0$ is associated with the unbroken phase, $\bar{z} > 0$ with the broken phase, and the boundary wall extends over $0 < \bar{z} < L_{w}$, where $L_{w}$ is the wall thickness.

    When $\bar{z} > 0$, the relaxation term $\bar{\Gamma}$ is non-zero and constant.
	
	\item For simplicity, assume that the source behavior typical of a simple step function: $\bar S$ nonzero and constant for $0 < \bar{z} < L_{w}$.

\end{itemize}
Then, the solution to Eq. (\ref{A7}) can be written as
\begin{eqnarray}
H(\bar{z}) = {\cal A} \,  e^{ v_{w}\bar{z}/\bar{D}} \ ,
\end{eqnarray}
with
\begin{eqnarray}
{\cal A} = k_H \, L_w \, \displaystyle\sqrt{\frac{\bar{\Gamma}}{\bar{D}}} \
\frac{ S_{\tilde{H}}^{CP} + S_{\tilde{t}}^{CP}
}{
\Gamma_h + \Gamma_{m}} \ ,
\end{eqnarray}
$ S_{\tilde{H}}^{CP}$ and $S_{\tilde{t}}^{CP}$ are respectively the Higgsino CP-violating source and the Quark CP-violating source.

Finally, by separating the contributions of the Higgsino CP-violating source and the Quark CP-violating source in Eq. (\ref{A3}), we can obtain:
\begin{eqnarray}
	&&Y_B=F_1\sin \theta_\mu+F_2\sin (\theta_\mu+\theta_T)\label{F11}.
\end{eqnarray}
The first term that contains $F_{1}$ stems from the Higgsino source, while the $F_{2}$ term arises from the squark source. And $\theta_\mu$ and $\theta_T$ are the corresponding phases.

\section{Constants $C_{abc}^{L,R}$ appeared in our calculation and Form factors.\label{C}}
\subsection{Constants $C_{abc}^{L,R}$}

\begin{eqnarray}
&&C_{\bar e_i \tilde v_k \tilde {\chi}_j^-}^L=U_{j,2}^*\sum_{b=1}^3 Z_{kb}^{V,*}\sum_{a=1}^3 U_{R,ia}^{e,*} Y_{e,ab}\ ,\nonumber\\
&&C_{\bar e_i \tilde v_k \tilde {\chi}_j^-}^R=-g_2\sum_{a=1}^3 Z_{ka}^{V,*} U_{L,ia}^e V_{j1}\ ,\\
&&C_{\tilde {\chi}_j^+ \tilde v_k^* e_j}^L=-g_2 V_{i1}^* \sum_{a=1}^3 U_{L,ja}^{e,*} Z_{ka}^V\ ,\nonumber\\
&&C_{\tilde {\chi}_j^+ \tilde v_k^* e_j}^R=\sum_{b=1}^3\sum_{a=1}^3 Y_{e,ab}^* U_{R,ja}^{e} Z_{kb}^{V} U_{i2}\ ,\\
&&C_{\tilde {\chi}_i^0 \tilde e_k^* e_j}^L=\frac{1}{2}(-2N_{i3}^*\sum_{b=1}^3U_{L,jb}^{e,*}\sum_{a=1}^3Y_{e,ab} Z_{k3+a}^{E}+\sqrt{2}g_1N_{i1}^*\sum_{a=1}^3U_{L,ja}^{e,*}Z_{ka}^{E}\nonumber\\
&&\qquad\qquad\;+\sqrt{2}g_2N_{i2}^*\sum_{a=1}^3U_{L,ja}^{e,*}Z_{ka}^{E})\ ,\nonumber\\
&&C_{\tilde {\chi}_i^0 \tilde e_k^* e_j}^R=-\sqrt{2}g_1\sum_{a=1}^3Z_{k3+a}^{E}U_{R,ja}^{e}N_{i1}-\sum_{b=1}^3\sum_{a=1}^3Y_{e,ab}^*U_{R,ja}^{e}Z_{kb}^{E}N_{i3}\ ,\\
&&C_{\bar e_i \tilde e_k \tilde {\chi}_j^0}^L=-2N_{j3}^*\sum_{b=1}^3Z_{kb}^{E,*}\sum_{a=1}^3U_{R,ia}^{e,*}Y_{e,ab}-\sqrt{2}g_1N_{j1}^*\sum_{a=1}^3Z_{k3+a}^{E,*}U_{R,ia}^{e,*}\ ,\nonumber\\
&&C_{\bar e_i \tilde e_k \tilde {\chi}_j^0}^R=\frac{1}{2}(-2\sum_{b=1}^3\sum_{a=1}^3Y_{e,ab}^{e,*}Z_{k3+a}^{E,*}U_{L,ib}^eN_{j3}+\sqrt{2}\sum_{a=1}^3Z_{ka}^{E,*}U_{L,ia}^e(g_1N_{j1}+g_2N_j2))\ ,\\
&&C_{\bar{\tilde g}\tilde q_i q_j}^L=-\sqrt2 g_3 Z^{\tilde q}_{i,j} e^{i\theta_3},\;\;
C_{\bar{\tilde g}\tilde q_i q_j}^R=\sqrt2 g_3 Z^{\tilde q}_{i,j+3} e^{-i\theta_3},\nonumber\\
&&C_{\bar{q}_j\tilde q_i \tilde g}^L=\sqrt2 g_3 Z^{\tilde q *}_{i,j+3} e^{i\theta_3},\;\;
C_{\bar{q}_j\tilde q_i \tilde g}^R=-\sqrt2 g_3 Z^{\tilde q *}_{i,j} e^{-i\theta_3},\\
&&C_{\bar{\chi}_k^0 \tilde u_i u_j}^L=-\frac{1}{6}\Big(\sqrt2 (g_1 Z^{N*}_{k,1}+3 g_2 Z^{N*}_{k,2})Z^{\tilde u}_{i,j}+6Y_{u,j}Z^{N*}_{k,4}Z^{\tilde u}_{i,j+3}\Big),\nonumber\\
&&C_{\bar{\chi}_k^0 \tilde u_i u_j}^R=-\frac{1}{6}\Big(\sqrt2(3g_2Z^{N}_{k,2} +g_1 Z^{N}_{k,1})Z^{\tilde u*}_{i,j}+6Y_{u,j}^*Z^{N}_{k,4}Z^{\tilde u*}_{i,j+3}\Big),\\
&&C_{\bar{\chi}_k^0 \tilde d_i d_j}^L=-\frac{1}{6}\Big(\sqrt2 (g_1 Z^{N*}_{k,1}-3 g_2 Z^{N*}_{k,2})Z^{\tilde d}_{i,j}+6Y_{d,j}Z^{N*}_{k,3}Z^{\tilde d}_{i,j+3}\Big),\nonumber\\
&&C_{\bar{\chi}_k^0 \tilde d_i d_j}^R=-\frac{1}{6}\Big(\sqrt2(-3g_2Z^{N}_{k,2} +g_1 Z^{N}_{k,1})Z^{\tilde d*}_{i,j}+6Y_{d,j}^*Z^{N}_{k,3}Z^{\tilde d*}_{i,j+3}\Big),\\
&&C_{\bar d_jd \tilde u_i {\chi}_k^-}^L=U_{k,2}^*\sum_{a=1}^3Y_{d,a}Z^{CKM*}_{a,j} Z^{\tilde u*}_{i,a},\nonumber\\
&&C_{\bar d_jd \tilde u_i {\chi}_k^-}^R=\sum_{a=1}^3 \Big(-g_2 V_{k,1}Z^{CKM}_{a,j} Z^{\tilde u*}_{i,a}+V_{k,2} Y_{u,a}^*Z^{CKM}_{a,j} Z^{\tilde u*}_{i,a+3}\Big).
\end{eqnarray}

\subsection{Form factors}
	
Defining ${x_i} = \frac{{m_i^2}}{{m_W^2}}$, we can find the form factors:
\begin{eqnarray}
	&&I_1(x)=\frac{1}{2(x-1)^2}(1+x+\frac{2x}{x-1}\ln x) \nonumber\\
	&&I_2(x)=\frac{1}{6(x-1)^2}(10x-26-\frac{2x-18}{x-1}\ln x) \nonumber\\
	&&I_3(x)=\frac{1}{2(x-1)^2}(3-x+\frac{2}{x-1}\ln x) \Big]\:,\\
	&&{I_1}(\textit{x}_1 , x_2 ) = \frac{1}{{16{\pi ^2}}}\Big[ \frac{{1 + \ln {x_2}}}{{({x_2} - {x_1})}} + \frac{{{x_1}\ln {x_1}}-{{x_2}\ln {x_2}}}{{{{({x_2} - {x_1})}^2}}} \nonumber\\
	&&{I_2}(\textit{x}_1 , x_2 ) = \frac{1}{{16{\pi ^2}}}\Big[ - \frac{{1 + \ln {x_1}}}{{({x_2} - {x_1})}} - \frac{{{x_1}\ln {x_1}}-{{x_2}\ln {x_2}}}{{{{({x_2} - {x_1})}^2}}} \nonumber\\
	&&{I_3}(\textit{x}_1 , x_2 ) = \frac{1}{{32{\pi ^2}}}\Big[  \frac{{3 + 2\ln {x_2}}}{{({x_2} - {x_1})}} - \frac{{2{x_2} + 4{x_2}\ln {x_2}}}{{{{({x_2} - {x_1})}^2}}} -\frac{{2x_1^2\ln {x_1}}}{{{{({x_2} - {x_1})}^3}}} \nonumber\\
	&&\qquad\qquad\quad\; + \: \frac{{2x_2^2\ln {x_2}}}{{{{({x_2} - {x_1})}^3}}}\nonumber\\
	&&{I_4}(\textit{x}_1 , x_2 ) = \frac{1}{{96{\pi ^2}}} \Big[ \frac{{11 + 6\ln {x_2}}}{{({x_2} - {x_1})}}- \frac{{15{x_2} + 18{x_2}\ln {x_2}}}{{{{({x_2} - {x_1})}^2}}} + \frac{{6x_2^2 + 18x_2^2\ln {x_2}}}{{{{({x_2} - {x_1})}^3}}}  \nonumber\\
	&&\qquad\qquad\quad\; + \: \frac{{6x_1^3\ln {x_1}}-{6x_2^3\ln {x_2}}}{{{{({x_2} - {x_1})}^4}}}  \Big]\:,\\
	&&J(x,y,z)=\ln x-\frac{y\ln y-z\ln z}{y-z} \Big]\:.\
\end{eqnarray}

\end{document}